# Coordinated Multi-Valve Disturbance-Rejection Pressure Control for High-Altitude Test Stands via Exterior Penalty Functions

**Zhang Louyue[a], Li Xin[b], Zhai Chao[b], Shi Duoqi[a], Zhang Hehong[c], Dan Zhihong[d], Wang Xi[a], Liu Jiashuai[e], Xiao Gaoxi[f]**

[a] *School of Energy and Power Engineering, Beihang University, Beijing 100191, China*

[b] *School of Automation, China University of Geosciences, Wuhan, 430074, China*

[c] *College of Computer and Data Science, Fuzhou University, Fuzhou, 350108, China*

[d] *Science and Technology on Altitude Simulation Laboratory, Sichuan Gas Turbine Establishment, Aero Engine Corporation of China, Mianyang, Sichuan, 621000, China*

[e] *Institute for Aero Engine, Tsinghua University, Beijing 100084, China*

[f] *School of Electrical and Electronic Engineering, Nanyang Technological University, 639798, Singapore*

**Abstract**: High-altitude simulation test benches for aero engines employ multi-chamber, multi-valve intake systems that demand effective decoupling and strong disturbance rejection during transient tests. This paper proposes a coordinated active disturbance rejection control (ADRC) scheme based on external penalty functions. The chamber-pressure safety limit is formulated as an inequality-constrained optimization problem, and an exponential penalty together with a gradient-based algorithm is designed for dynamic constraint relaxation, with guaranteed global convergence. A coordination term is then integrated into a distributed ADRC framework to yield a multi-valve coordinated ADRC controller, whose asymptotic stability is established via Lyapunov theory. Hardware-in-the-loop simulations using MATLAB/Simulink and a PLC demonstrate that, under ±3 kPa pressure constraints, chamber V2's maximum error is 1.782 kPa (77.1% lower than PID control), and under a 180 $kg/s^2$ flow-rate disturbance, valve oscillations decrease from ±27% to ±5%. These results confirm the superior disturbance-rejection and decoupling performance of the proposed method.

## Introduction

High-altitude simulation testing represents the most efficient means of exploring, improving, and troubleshooting aero-engine design and performance characteristics, as well as of reproducing and eliminating in-flight faults. A critical component of such testing is the accurate simulation of the engine's airborne operating environment. Among the subsystems of a flight-environment simulator, the inlet-pressure control system is one of the core units, and it provides a stable pressure environment at the engine intake[1]-[3].

Large-flow aero-engines are currently the focus of high-altitude transient thrust, inertial start, and acceleration/deceleration experiments conducted on altitude test stands. These transient tests first establish steady conditions corresponding to a given flight altitude and Mach number via the environmental-simulation control system. Then, by rapidly moving the throttle lever (within $\leqslant$ 1.0 s), the engine is driven from one operating point to another. Each test typically lasts for only a few seconds, during which the engine's air-mass flow varies drastically (with maximum magnitudes up to 220% and peak rates exceeding 150 kg/s$^2$). The control system must ensure excellent dynamic performance and stability—fast response, minimal overshoot, and short settling time—under the large, rapid disturbances induced by aggressive throttle motions and engine acceleration/deceleration[4][5].

To accommodate such dramatic flow variation and disturbance rejection requirements, high-altitude test facilities employ a large, complex inlet-air system for high-flow engines. This system comprises multiple interconnected volumes whose parameters are strongly coupled and mutually interfering. To achieve rapid redistribution and allocation of airflow within a short time, several control valves are installed between adjacent chambers for coordinated regulation[6].

However, the multi-chamber coupling and multi-valve coordination present significant challenges for the design of control system. First, strong coupling among chambers means any valve movement generates pressure waves that trigger cascading disturbances in adjacent volumes. This makes traditional single-loop control strategies inadequate for decoupling multiple variables. Second, the system exhibits pronounced nonlinearity and time-varying behavior. During engine transients, valves swing widely and their flow coefficients display hysteresis. The volumetric effects of multiple chambers introduce high-order dynamic delays, causing the model parameters to vary sharply with engine operating conditions. Furthermore, stochastic disturbances—such as combustion instability and mechanical vibrations of the simulation equipment—compound the difficulty of designing a robust controller. These challenges call for advanced control algorithms with strong decoupling capability, multi-actuator coordination, and adaptive disturbance-rejection functions[7][8].

Researchers have undertaken extensive studies on coordinated control of multiple actuators. Reference [9] applies a linear-matrix-inequality (LMI)-based method to allocate valve flows and achieve coordination. Reference [10] combines open-loop and closed-loop control to decouple and synchronize multiple valves. Reference [11] proposes a multimodal switching framework—integrating adaptive sliding-mode control (ASMC) with a disturbance observer (DOB)—to coordinate actuators in a telescope-scanning imaging system. This approach dynamically switches between compensation and reset modes, effectively resolving image-motion degradation due to mirror rotation, and uses an adaptive reaching law to suppress chattering in sliding-mode control. Although these methods demonstrate favorable results in their respective simulation environments, they generally require high linearity and limited uncertainty in the plant model—conditions not met by the highly nonlinear, strongly disturbed, large-scale MIMO inlet-pressure system.

The inlet-environment pressure control problem can be recast as a constrained optimization problem. Among the solution techniques, penalty-function methods transform constrained problems into unconstrained ones by incorporating constraint violations into the objective, thereby simplifying the solution process[12][13]. Penalty functions have been widely applied in aerospace control applications. In fixed-wing low-altitude flight[14], they handle aerodynamic constraints, attitude limits, and dynamic environment restrictions to enhance safety, stability, and efficiency. For electric flying-wing UAVs, adaptive penalty-function methods optimize flight parameters under payload and endurance constraints, improving operational safety[15]. In UAV obstacle-avoidance research, penalty-function-defined safe regions boost flight safety in uncertain environments. Similarly, the inlet-pressure control task seeks to keep each chamber's pressure within a prescribed error bound while minimizing simulation deviation, which can be cast as a constrained optimization problem and solved via a penalty-function approach.

Cooperative control denotes the process by which multiple subsystems work together toward a shared objective[16]-[18]. In current multi-chamber, multi-valve (MCMV) control systems, the chamber pressure is controlled independently: when a chamber has multiple regulating valves, individual valves are governed by separate single-input single-output (SISO) active disturbance rejection controllers (ADRC). Although ADRC's intrinsic decoupling endows the system with basic coordination[19]-[21], existing methods suffer from three major drawbacks. First, the sluggish dynamic response cannot accommodate rapid flow fluctuations, causing significant pressure overshoot. Second, without an active coordination mechanism to handle strong coupling among chambers and valves, the system—relying solely on ADRC's passive decoupling—is prone to pressure oscillations under complex disturbances. Third, the control architecture grows linearly with valve count, leading to an exponential rise in tuning complexity and severely restricting practical deployment. Therefore, deeply integrating ADRC advantages with cooperative control theory to develop a multivariable coordination algorithm for strongly coupled scenarios is the key to improving overall system performance.

Motivated by the foregoing, this paper proposes a penalty function based coordinated ADRC for multi-chamber inlet-pressure regulation. The main contributions are given as follows:

1. We propose a cooperative optimization algorithm based on penalty functions for pressure control in MVMC systems. The algorithm maintains the simulation deviation at the engine inlet within predefined limits. It also improves the stability of chamber pressure. The convergence of the algorithm is theoretically guaranteed.

2. We develop a coordinated ADRC scheme to address the inefficiency of

independent control loops, improving control efficiency and robustness. Closed-loop stability is rigorously analyzed.

3. We implement a hardware-in-the-loop (HIL) simulation to compare the proposed method with classical PID control, demonstrating its feasibility and superiority.

The remainder of this paper is organized as follows. Section 1 presents the multi-chamber pressure-control problem, system-structure characteristics, and control challenges. Section 2 derives the coupled pressure–temperature differential equations for the chambers and the nonlinear dynamic model of the control valves. Section 3 formulates the multi-objective cooperative optimization problem via exterior penalty functions and establishes convergence proofs. Section 4 designs the coordinated ADRC with disturbance compensation and performs closed-loop stability analysis. Section 5 validates the proposed algorithm on a hardware-in-the-loop platform and compares results with traditional PID control. Finally, Section 6 concludes the study and discusses future research directions.

# 1 Problem Statement

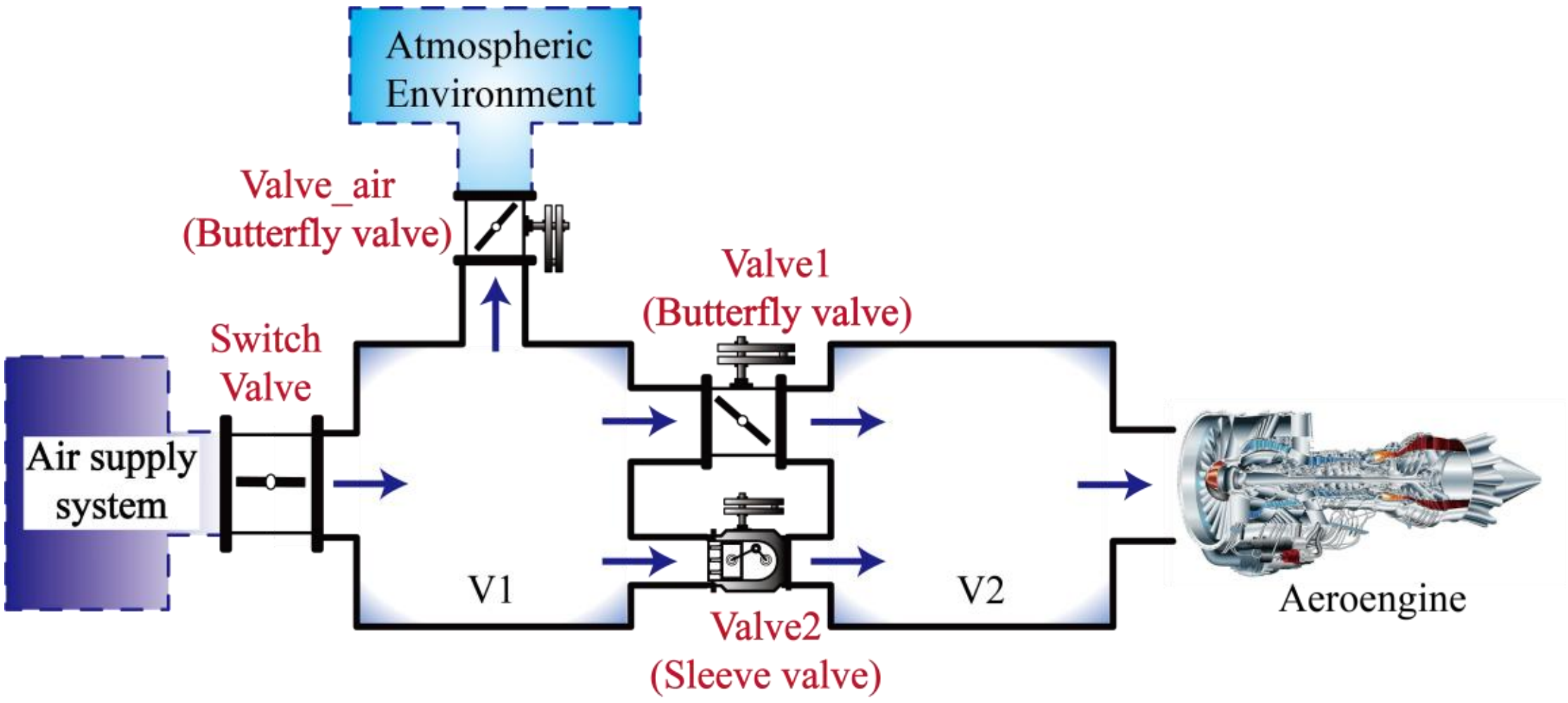

Fig. 1 Schematic diagram of system structure

Figure. 1 shows a schematic of the MVMC intake pressure control system. The system comprises two chambers: V1 and V2. The pressure in V2, representing the engine inlet pressure, is regulated by a butterfly valve (Valve1) and a sleeve valve (Valve2). The pressure in V1, which serves as the upstream boundary for V2, is controlled by a butterfly valve (Valve_air). A supply system provides constant inlet pressure and temperature to V1, and Valve_air exhausts to the ambient environment.

In testing, to ensure that V2 pressure tracks its setpoint and rejects disturbances from engine bleed flow variations, the control valves must operate cooperatively to match flow across all sections. However, when the bleed flow changes rapidly and significantly (e.g., a flow-rate change up to 150 $kg/s^2$), a single closed-loop control framework cannot guarantee effective valve coordination. Simulations exhibit valve saturation, pressure oscillations, and control lag, resulting in regulation failure that does not meet the technical requirements of current tests. Engine test specifications require that the simulated V2 pressure never deviate by more than 3 kPa from its preset value at any point during the intake-environment simulation. This limit ensures high-quality replication under both steady and transient conditions. Additionally, to maintain safe operation of the supply unit throughout the test, the pressure in V1 should remain within ±5 kPa of its setpoint.

# 2 Inlet-air System Model

## 2.1 Chamber Temperature and Pressure Model

As shown in Fig. 1, the chamber V1 has one inlet and three outlets, while the chamber V2 has two inlets and one outlet (see Fig. 2). In the diagram, $\dot{m}$ 、 $C$ 、 $T$ 、 $P$ 、 $V$ 、 $\dot{Q}$ denote the mass flow rate, mean flow velocity, temperature, pressure, chamber volume, and heat transfer rate between the chamber and its surroundings, respectively. The subscript “in” refers to the main supply manifold parameters; “air” to the flow parameters at Valve_air; “V1” and “V2” to those at Valve1 and Valve2; and “1” and “2” to the gas parameters in chambers V1 and V2, respectively.

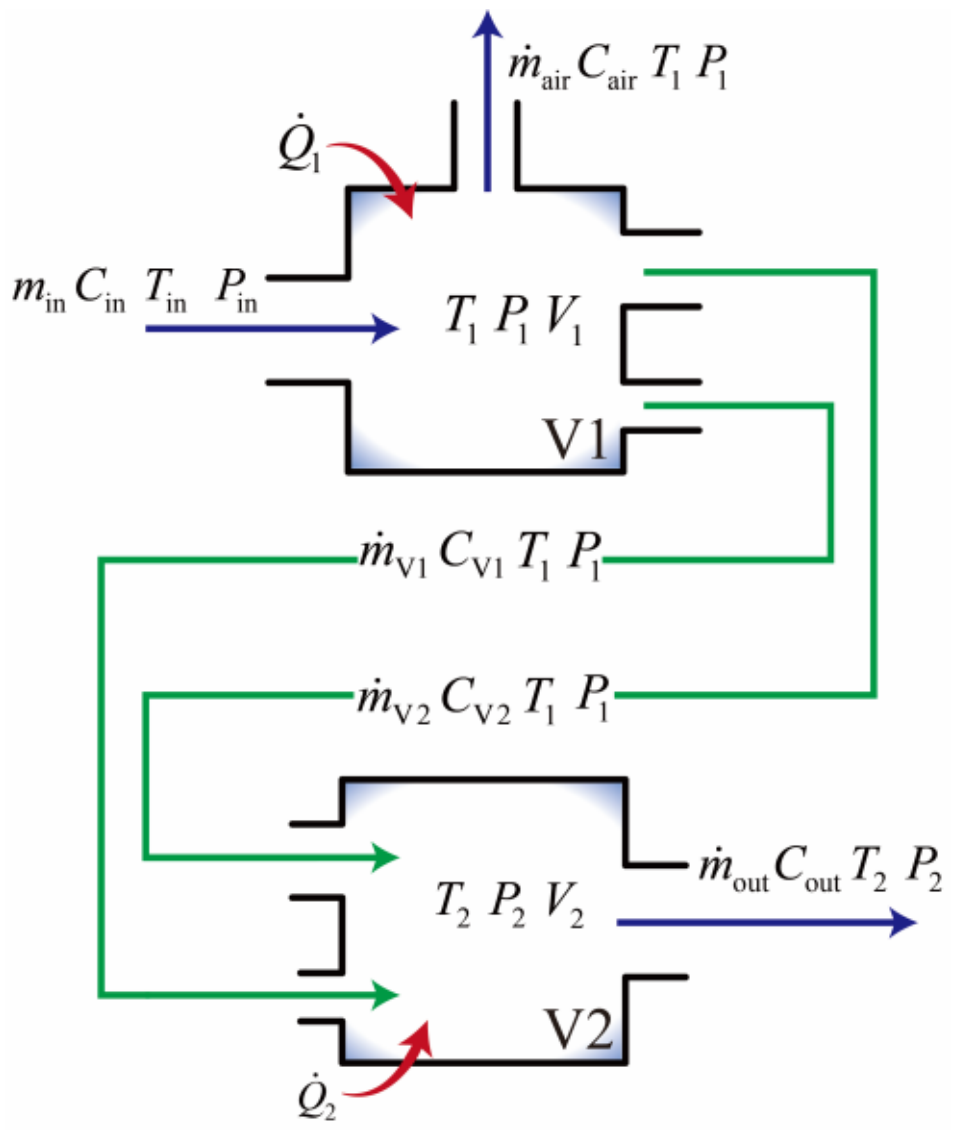

Fig. 2 Chamber structure diagram

Since the chamber volumes dominate the rates of pressure and temperature change and the connecting pipes are short, pipeline dynamics are neglected. Accordingly, a lumped-parameter approach is used to model pressure and temperature dynamics in the stabilizing chambers. Defining control volumes 1 and 2 by the chamber walls and inlet/outlet boundaries and neglecting changes in potential energy, we derive the differential equations for pressure and temperature in V1 and V2 from the continuity equation, the energy equation, and the ideal-gas law:

$$
\begin{aligned}
\frac{dP_1}{dt} &= \frac{RT_1}{V_1}\left[\dot{m}_{in} - \left(\dot{m}_{air} + \dot{m}_1 + \dot{m}_2\right)\right] \\
&+ \frac{R}{V_1(C_p - R)}\left[\dot{Q}_1 + \dot{m}_{in}\left(h_{in} + \frac{1}{2}C_{in}^2\right) - \dot{m}_{air}\left(h_{air} + \frac{1}{2}C_{air}^2\right)\right. \\
&\left. - \dot{m}_1\left(h_1 + \frac{1}{2}C_1^2\right) - \dot{m}_2\left(h_2 + \frac{1}{2}C_2^2\right) - C_p T_1\left[\dot{m}_{in} - \left(\dot{m}_{air} + \dot{m}_1 + \dot{m}_2\right)\right]\right]
\end{aligned}
\tag{1}
$$

$$
\begin{aligned}
\frac{dT_1}{dt} &= \frac{RT_1}{P_1 V_1(C_p - R)}\left[\dot{Q}_1 + \dot{m}_{in}\left(h_{in} + \frac{1}{2}C_{in}^2\right) - \dot{m}_{air}\left(h_{air} + \frac{1}{2}C_{air}^2\right)\right. \\
&\left. - \dot{m}_1\left(h_1 + \frac{1}{2}C_1^2\right) - \dot{m}_2\left(h_2 + \frac{1}{2}C_2^2\right) - C_p T_1\left[\dot{m}_{in} - \left(\dot{m}_{air} + \dot{m}_1 + \dot{m}_2\right)\right]\right]
\end{aligned}
\tag{2}
$$

$$
\begin{aligned}
\frac{dP_2}{dt} &= \frac{RT_2}{V_2}\left(\dot{m}_1 + \dot{m}_2 - \dot{m}_{out}\right) + \frac{R}{V_2(C_p - R)}\left[\dot{Q}_1 + \dot{m}_1\left(h_1 + \frac{1}{2}C_1^2\right)\right. \\
&\left. + \dot{m}_2\left(h_2 + \frac{1}{2}C_2^2\right) - \dot{m}_{out}\left(h_{out} + \frac{1}{2}C_{out}^2\right) - C_p T_2\left(\dot{m}_1 + \dot{m}_2 - \dot{m}_{out}\right)\right]
\end{aligned}
\tag{3}
$$

$$\frac{dT_2}{dt}=\frac{RT_2}{P_2V_2(C_p-R)}\left[\dot{Q}_1+\dot{m}_1\left(h_1+\frac{1}{2}C_1^2\right)+\dot{m}_2\left(h_2+\frac{1}{2}C_2^2\right)\right.$$
$$\left.-\dot{m}_{out}\left(h_{out}+\frac{1}{2}C_{out}^2\right)-C_pT_2\left(\dot{m}_1+\dot{m}_2-\dot{m}_{out}\right)\right] \quad (4)$$

In these expressions, $P_1$、$T_1$、$V_1$ and $T_2$、$P_2$、$V_2$ denote the pressure, temperature, and volume of chambers V1 and V2, respectively. The terms $h_{out}$、$h_{in}$、$h_1$、$h_2$、$h_{air}$ are the specific enthalpies of the airflow at five sections: the outlet of V2, the inlet of V1, through Valve1, Valve2, and Valve_air, respectively. $\dot{Q}_1$ and $\dot{Q}_2$ are the net rates of heat transfer between chambers V1 and V2 and the surroundings, respectively. Finally, $C_P$ is the specific heat at constant pressure, and *R* is the gas constant.

## 2.2 Control Valve Model

The intake pressure control system employs large-diameter butterfly and sleeve valves as flow-regulating elements. Both valve types are actuated by a hydraulic-servo closed-loop control system. Consequently, the valve model includes a flow-characteristic model and a dynamic-characteristic model, as shown in Fig. 3. In this figure, $VP_{\text{cmd}}$ denotes the valve-opening command, $A_{\text{act}}$ denotes the actual equivalent flow area, and $\dot{m}_{\text{out}}$ denotes the actual flow rate through the valve.

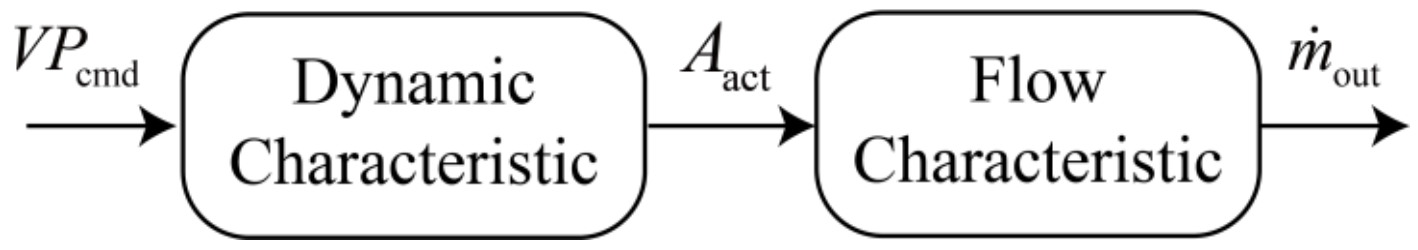

Fig. 3 Diagram of control valve model

### 2.2.1 Valve Flow Characterization Models

The theoretical flow equations for butterfly and sleeve valves are complex and not convenient for practical use. Therefore, the flow coefficient method is adopted for modeling, and a generalized flow equation is used to characterize their flow characteristics [ ], as follows:

$$\dot{m} = \varphi A p \sqrt{\frac{2}{RT}} \tag{5}$$

In this equation, $\dot{m}$ denotes the gas mass flow rate through the control valve; $\varphi$ is the flow coefficient obtained by fitting experimental data; $\rho$ is the chamber gas density; $A$ is the valve's equivalent flow area; $p$ and $T$ are the upstream gas pressure and temperature, respectively. The coefficient $\varphi$ depends on the upstream pressure $p$, downstream pressure $p_a$, and equivalent flow area $A$, as given in below.

$$\varphi = f\left(p, p_a, A\right) \tag{6}$$

**2.2.2 Dynamic Model of Valve Actuation Characteristics**

The valve plate is driven by a hydraulic cylinder through a linkage mechanism, and the spool displacement is controlled by a closed-loop negative-feedback system. With optimized controller parameters, the dynamic response of this displacement loop can be approximated as a first-order inertial process. Due to clearances between the valve plate and linkages, hysteresis arises during actuation. This is represented by a nonlinear function $f_{VP}$. The valve's dynamic characteristic model is shown in Fig. 4, where $\tau$ is the time constant.

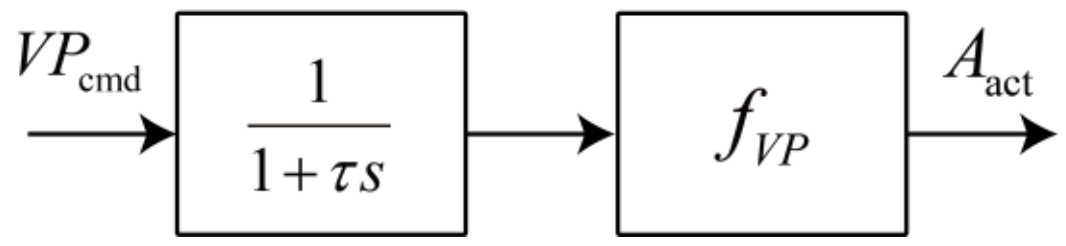

Fig. 4 Schematic Diagram of the Valve Dynamic Characteristic Model

# 3. Design of Pressure Coordination Control Algorithm Based on Penalty Function

## 3.1 Description of Optimization Problem

Based on the problem statement in Section 1, the control objectives of the intake pressure control system are to ensure that the pressures in chambers V1 and V2 track their setpoints or reject disturbances. The constraints impose bounds on the deviations between actual and preset pressures in V1 and V2. Therefore, the problem is cast as an optimization problem with inequality constraints:

$$\begin{aligned} &\min f\left(P_1,P_2\right) \\ &s.\,t.\; g_i\left(P_1,P_2\right)\le 0 \end{aligned} \tag{7}$$

where, $f,g_i: R^n \to R, i\in\{1,2\}$ is a continuously differentiable function. $S=\{P_1,P_2\in R^n \mid g_i\left(P_1,P_2\right)\le 0, i\in\{1,2\}\}$ denotes the feasible domain of Problem (7).

## 3.2 Construction of Penalty Function

The penalty-function method converts a constrained optimization problem into an unconstrained one. It is a key approach for solving optimization problems with inequality constraints. Specifically, with target setpoints $P_{1\mathrm{set}}$ and $P_{2\mathrm{set}}$, the objective function $f\left(P_1,P_2\right)$ in the optimization problem is defined as:

$$f\left(P_1,P_2\right)=\left(P_1-P_{1\mathrm{set}}\right)^2+\left(P_2-P_{2\mathrm{set}}\right)^2 \tag{8}$$

In addition, the valid simulation pressure range for the two chambers, $g_i\left(P_1,P_2\right)$, can be expressed as:

$$\begin{aligned} g_1\left(P_1,P_2\right)&=\left(P_1-P_{1\mathrm{set}}\right)^2-\varepsilon_1^2 \\ g_2\left(P_1,P_2\right)&=\left(P_2-P_{2\mathrm{set}}\right)^2-\varepsilon_2^2 \end{aligned} \tag{9}$$

where $\varepsilon_1$ and $\varepsilon_2$ represent the maximum allowable deviations from the setpoint pressures in chambers V1 and V2, respectively. Based on the above inequality constraints, the augmented objective function is defined as follows:

$$L\left(P_1,P_2,\mu\right)=f\left(P_1,P_2\right)+\gamma\alpha\left(P_1,P_2,\eta\right) \tag{10}$$

where $\gamma>0$ is the penalty factor, $\alpha\left(P_1,P_2,\eta\right)$ is the penalty function, and $\eta=[\mu,\sigma]>0$ is the tuning parameter. Based on the characteristics of the exponential function, the penalty function is designed as follows:

$$\alpha\left(P_1,P_2,\eta\right)=\sum_{i=1}^{2}\left(e^{\max\{0,\eta g_i\left(P_1,P_2\right)\}}-1\right)^2 \tag{11}$$

## 3.3 Design of Coordinated Optimization Control Algorithm

Based on the control objectives and the penalty function, the constrained problem can be transformed into:

$$\min_{P_1,P_2} L\left(P_1,P_2,\gamma\right) \tag{12}$$

When the system reaches optimality, the augmented objective function attains its optimum values $\left\{P_1^*,P_2^*\right\}=\arg\min L\left(P_1,P_2,\gamma\right)$. It ensures that both chamber pressures converge rapidly to their setpoints and remain within the prescribed error bounds throughout regulation. The algorithm is as elaborated follows:

**Step 1: Parameter Initialization**

Initialize the state variables $P_1, P_2$, and choose the initial penalty factor $\gamma_0$, tuning coefficient $\eta_0=[\mu_0,\sigma_0]$, and gradient learning rate $l$.

**Step 2: Iterative Process**

(a) Construct the augmented objective function $L\left(P_1,P_2,\gamma\right)$ by adding penalty terms according to the constraints, then compute its gradient $\nabla L\left(P_1,P_2,\gamma\right)=\left[\frac{\partial L\left(P_1,P_2,\gamma\right)}{\partial P_1},\frac{\partial L\left(P_1,P_2,\gamma\right)}{\partial P_2}\right]^{\mathrm{T}}$.

(b) Use a gradient-based optimization method to solve the penalized problem and obtain the solution at iteration $k$.

$$\min_{p_1,p_2} L_k\left(P_{1,k},P_{2,k},\gamma_k\right) \tag{13}$$

(c) Update the state variables.

$$\begin{aligned} P_{1,k+1} &= P_{1,k} - l\nabla L_k(P_{1,k},P_{2,k},\gamma_k)e_1 \\ P_{2,k+1} &= P_{2,k} - l\nabla L_k(P_{1,k},P_{2,k},\gamma_k)e_2 \end{aligned} \tag{14}$$

(d) Update the penalty coefficient $\gamma_{k+1}=\omega_1\gamma_k$.

**Step 3: Termination Condition**

Stop the iteration when either the solution $P_1,P_2$ converges within the required tolerance or the maximum number of iterations is reached. Additionally, if $\gamma_k \alpha(P_{1,k+1}, P_{2,k+1}, \eta_0) < \xi, \xi > 0$, terminate the algorithm.

## 3.4 Convergence Analysis

The existence of a feasible point is a prerequisite for solvability of the optimization problem. This gives rise to the following lemma.

**Lemma 3.1**

For any $\gamma > 0, \eta = [\mu, \sigma] > 0$ and $f(P_1, P_2) = (P_1 - P_{1\text{set}})^2 + (P_2 - P_{2\text{set}})^2$, suppose there exists $(P_1^*, P_2^*)$ such that

$$\theta(\gamma) = \inf\{f(P_1, P_2) + \gamma\alpha(P_1, P_2, \eta)\} = L(P_1^*, P_2^*, \gamma) \tag{15}$$

Then:

(1) $\inf\{f(P_1, P_2) | \ g_i(P_1, P_2) \le 0\} \ge \sup \theta(\gamma)$

(2) The penalty function $\alpha(P_1, P_2, \eta)$ is a nonincreasing function of $\gamma$, whereas both $f(P_1, P_2)$ and $\theta(\gamma)$ are nondecreasing in $\gamma$.

**Proof.**

Consider a feasible point $(P_1, P_2)$ in the constraint set

$$S = \{P_1, P_2 \in R^n | \ g_i(P_1, P_2) \le 0, i \in \{1,2\}\}, \alpha(P_1, P_2, \eta) = \sum_{i=1}^{2}\left(e^{\max\{0, \eta g_i(P_1, P_2)\}} - 1\right)^2 = 0,$$

(1) $f(P_1, P_2) = f(P_1, P_2) + \gamma\alpha(P_1, P_2, \eta) \ge \inf\{f(P_1, P_2) + \gamma\alpha(P_1, P_2, \eta)\} = \theta(\gamma),$

Therefore, $\inf\{f(P_1, P_2) | \ g_i(P_1, P_2) \le 0\} \ge \sup \theta(\gamma).$

(2) By the properties of $\theta(\gamma)$, assume $\gamma > \gamma'$. Then we have

$$f(P_1', P_2') + \gamma\alpha(P_1', P_2', \eta) \ge f(P_1, P_2) + \gamma\alpha(P_1, P_2, \eta), \tag{16}$$

$$f(P_1, P_2) + \gamma'\alpha(P_1, P_2, \eta) \ge f(P_1', P_2') + \gamma'\alpha(P_1', P_2', \eta). \tag{17}$$

Adding these two inequalities yields $(\gamma - \gamma')(\alpha(P_1', P_2', \eta) - \alpha(P_1, P_2, \eta)) \ge 0,$ which implies that $\alpha(P_1, P_2, \eta)$ is nonincreasing in $\gamma$.

From inequality (17) we also obtain $f(P_1, P_2) - f(P_1', P_2') \ge 0$, hence $f(P_1, P_2)$ is nondecreasing in $\gamma$. Finally, inequality (17) further implies

$$f(P_1,P_2)+\gamma'\alpha(P_1,P_2,\eta)\geq f(P_1',P_2')+\gamma'\alpha(P_1',P_2',\eta)=L(P_1',P_2',\gamma')=\theta(\gamma'),$$
$$\theta(\gamma)+(\gamma'-\gamma)\alpha(P_1,P_2,\eta)\geq\theta(\gamma'),$$

establishing that $\theta(\gamma)$ is nonincreasing in $\gamma$.

This completes the proof.

As $\gamma$ grows, the solution of the penalized problem diverges, and the value of $L$ approaches the original objective function $f(P_1,P_2)$. This means that as the penalty for constraint violations becomes arbitrarily large, the solution increasingly approximates the global optimum of the original problem.

**Theorem 3.1**

Any limit point of a convergent subsequence is a global optimal solution, and $\lim_{\gamma\to+\infty} L(P_1,P_2,\gamma)=f(P_1,P_2).$

**Proof.**

The proof details include three parts.

**(a) Boundedness of the penalty term**

Let $\left(P_1^*,P_2^*\right)=\arg\min_{P_1,P_2} L\left(P_1,P_2,\gamma\right)$. By Lemma 3.1,

$$\begin{aligned}&\inf\{f(p_c)|\ g_i(p_c)\leq 0,h_i(p_c)-\gamma_i\leq 0\}\\&\geq\theta(\mu,\sigma)=f(p_c)+\mu\alpha(p_c)+\sigma\beta(p_c,\gamma_i)\\&\geq f(p_c^*)+\mu\alpha(p_c^*)+\sigma\beta(p_c^*,\gamma_i)>f(p_c).\end{aligned}$$

Since $\alpha(P_1,P_2,\eta)$ is nonincreasing in $\gamma$, it follows that $f(P_1,P_2)-f(P_1^*,P_2^*)\leq\gamma\alpha(P_1,P_2,\eta)\leq\xi.$ Hence $\alpha$ remains bounded as $\gamma$ grows.

**(b) Feasibility of the limit point**

As $\gamma\to+\infty$, both $\xi\to 0$ and $\alpha(P_1,P_2,\eta)\to 0$. For any convergent subsequence $\left\{P_{1,k},P_{2,k}\right\}$, Lemma 3.1 gives $\sup_{\gamma\geq 0}\theta(\gamma)\geq\theta(\gamma,k)=f(P_1,P_2)+\gamma_k\alpha(P_1,P_2,\eta)\geq f(P_1,P_2)$. Let $\left(P_1^*,P_2^*\right)$ be its limit of $\left\{P_{1,k},P_{2,k}\right\}$, then $\sup_{\gamma\geq 0}\theta(\gamma)\geq f(P_{1,k}^*,P_{2,k}^*)$. Since $\gamma\alpha(P_{1,k}^*,P_{2,k}^*,\eta)\to 0$, $P_1^*,P_2^*$ satisfies all original constraints of Problem (7).

**(c) Convergence of the augmented objective**

Again by Lemma 3.1, $\left(P_1^*, P_2^*\right)$ is an optimal solution and $\sup_{\gamma\geq 0}\theta(\gamma) = f(P_1^*, P_2^*)$.

As $\gamma \to +\infty$, $\gamma\alpha(P_1, P_2, \eta) = \theta(\gamma) - f(P_1, P_2) = f(P_1^*, P_2^*) - f(P_1^*, P_2^*) = 0$. Therefore, $\lim_{\gamma\to+\infty} L(P_1, P_2, \gamma) = f(P_1, P_2)$.

# 4 Design of Cooperative Controller

Figure 5 shows the overall control framework. The dual-chamber intake system employs three independent control valves to achieve coordinated pressure regulation. To guarantee that the pressures in both chambers satisfy the prescribed error bounds throughout the control period, a coordinated ADRC based on the penalty function optimization algorithm presented in Section 3 is used. During operation, the algorithm parameters are iteratively updated based on the real-time system state. The optimization problem is then repeatedly solved until the pressure control objectives are met and stable convergence is achieved. In Fig. 5, the blue region denotes the coordinated control algorithm; the green region, the primary controller; the yellow region, the decoupling module; the purple region, the simulation model; and the pink region, the extended state observer (ESO).

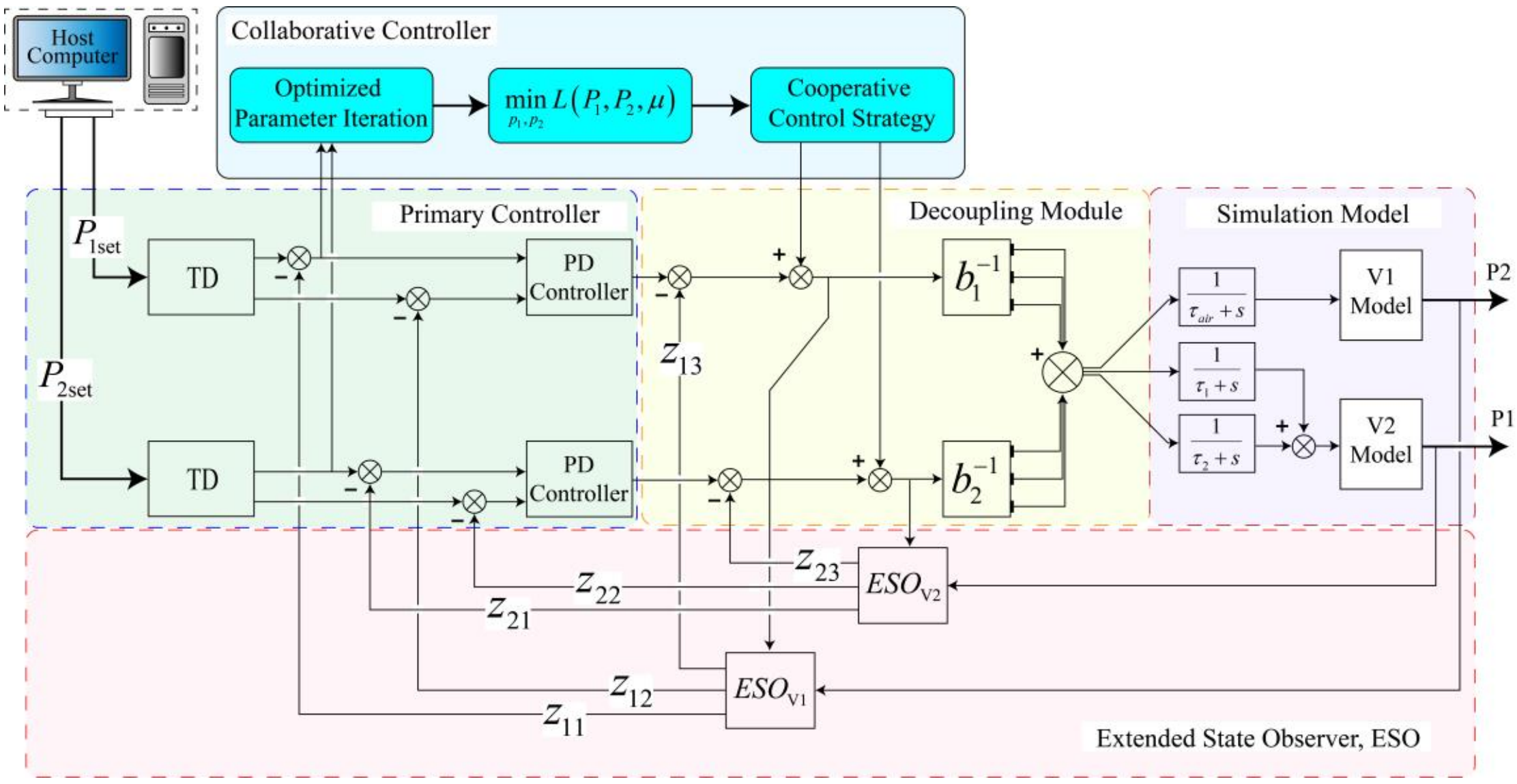

Fig. 5 Schematic Diagram of the Pressure Coordination Control System

## 4.1 Pressure Control Model of Dual-Chamber System

The dual-chamber system considered in this study features large chamber volumes and relatively slow internal airflow. Additionally, temperature variations are limited during flight environment simulation tests. Therefore, the effects of gas kinetic energy and heat transfer between the airflow and pipeline walls are neglected in the control model. Based on this, and using Equations (1) and (3), the simplified state-space equations of chamber pressures with respect to $\dot{m}_{air}$, $\dot{m}_1$, $\dot{m}_2$ are derived as follows:

$$\begin{bmatrix} \dot{P}_1 \\ \dot{P}_2 \end{bmatrix} = \begin{bmatrix} -\alpha_1 & -\alpha_2 & -\alpha_{air} \\ \beta_1 & \beta_2 & 0 \end{bmatrix} \begin{bmatrix} \dot{m}_1 \\ \dot{m}_2 \\ \dot{m}_{air} \end{bmatrix} + \begin{bmatrix} f_1(P_1,P_2) \\ f_2(P_1,P_2) \end{bmatrix} \tag{18}$$

where $\alpha_1 = \frac{R}{V_1}\left[T_1 - \frac{h_{in} - RT_1}{C_p - R}\right]$, $\alpha_2 = \frac{R}{V_1}\left[T_1 - \frac{h_2 - RT_1}{C_p - R}\right]$, $\alpha_{air} = \frac{R}{V_1}\left[T_1 - \frac{h_{air} - RT_1}{C_p - R}\right]$, $\beta_1 = \frac{R}{V_2}\left[a_{out} - \frac{h_1}{C_p - R}\right]$, $\beta_2 = \frac{R}{V_2}\left[a_{out} - \frac{h_2}{C_p - R}\right]$, $\alpha_{out} = T_2 - \frac{h_{out} - RT_2}{C_p - R}$. $f_1(P_1,P_2)$ and $f_2(P_1,P_2)$ represent the total disturbances in the two pressure control systems, including factors such as the neglected heat transfer between airflow and pipelines, the inlet mass flow rate $\dot{m}_{in}$ into Chamber 1, and the engine extraction flow rate $\dot{m}_{out}$ from Chamber 2.

## 4.2 Design of Coordinated ADRC Controller

Construct ESO for the chamber pressure control systems V1 and V2 based on (23) and (24):

$$\begin{cases} e_1 = z_{11} - y_1 \\ \dot{z}_{11} = z_{12} - \beta_{11} e_1 \\ \dot{z}_{12} = z_{13} - \beta_{12} e_1 + b_1 \left[VP_{1\mathrm{cmd}} \quad VP_{2\mathrm{cmd}} \quad VP_{3\mathrm{cmd}}\right]^T \\ \dot{z}_{13} = -\beta_{13} e_1 \end{cases} \tag{19}$$

$$\begin{cases} e_2 = z_{21} - y_2 \\ \dot{z}_{21} = z_{22} - \beta_{21} e_2 \\ \dot{z}_{22} = z_{23} - \beta_{22} e_2 + b_2 \begin{bmatrix} VP_{1\text{cmd}} & VP_{2\text{cmd}} & VP_{3\text{cmd}} \end{bmatrix}^T \\ \dot{z}_{23} = -\beta_{23} e_2 \end{cases} \tag{20}$$

where

$$z_1 = \begin{bmatrix} z_{11} \\ z_{12} \\ z_{13} \end{bmatrix}, \; z_2 = \begin{bmatrix} z_{21} \\ z_{22} \\ z_{23} \end{bmatrix}$$

are the state estimations for $P_1$, $P_2$, respectively, and $z_{13}$, $z_{23}$ denote the total disturbance estimations.

$$\beta_1 = \begin{bmatrix} \beta_{11} \\ \beta_{12} \\ \beta_{13} \end{bmatrix}, \; \beta_2 = \begin{bmatrix} \beta_{21} \\ \beta_{22} \\ \beta_{23} \end{bmatrix}$$

are the gain vectors for the two ESOs, respectively. The variable $e(t)$ represents the output state estimation error. Let $\omega_1$ and $\omega_2$ denote the observer bandwidths of the two ESOs. According to the bandwidth-based tuning method [ref], the gain vectors can be selected as follows:

$$\begin{gathered} \beta_{11} = 3\omega_1, \beta_{12} = 3\omega_1^2, \beta_{13} = \omega_1^3 \\ \beta_{21} = 3\omega_2, \beta_{22} = 3\omega_2^2, \beta_{23} = \omega_2^3 \end{gathered} \tag{21}$$

Assuming that the LESO estimates are accurate, the total disturbances acting on the system satisfy $z_{13} \to f_1(P_1, P_2)$, $z_{23} \to f_2(P_1, P_2)$, and hence the control inputs $U_1$ and $U_2$ in (23) and (24) can be written as

$$\begin{gathered} U_1 = b_1^{-1}\left(u_{c1} - f_1(P_1, P_2)\right) \\ U_2 = b_2^{-1}\left(u_{c2} - f_2(P_1, P_2)\right) \end{gathered}$$

Combining the optimization objective with a proportional–derivative (PD) controller yields the following cooperative control laws:

$$\begin{gathered} u_{c1} = k_{11}\left(P_{1\text{set}} - z_{11}\right) - k_{12}\left(\dot{P}_{1\text{set}} - z_{12}\right) - k_{13}\nabla L\left(P_1, P_2, \mu\right) \\ u_{c2} = k_{21}\left(P_{2\text{set}} - z_{21}\right) - k_{22}\left(\dot{P}_{2\text{set}} - z_{22}\right) - k_{23}\nabla L\left(P_1, P_2, \mu\right) \end{gathered} \tag{22}$$

where, $P_{1set}$ and $P_{2set}$ are the pressure setpoints for chambers V1 and V2, respectively; $\dot{P}_{1set}$ and $\dot{P}_{2set}$ are obtained via a tracking differentiator (TD)[23]; The feedback gain matrix is $K=[K_1,K_2]$ with $K_1=[k_{11},k_{12}]$, $K_2=[k_{21},k_{22}]$. $\nabla L(P_1,P_2,\mu)$ denotes the gradient of the cooperative optimization term, and $K_3=[k_{13},k_{23}]$ comprises the tuning parameters for this optimization control.

## 4.3 Stability Analysis

In this section, the stability of the controller designed for the aforementioned control strategy is established. First, by combining the system model, the controller, and the observer, the closed-loop system is obtained as follows:

$$\begin{cases} \dot{x}(t) = A_z x(t) + B_z U(t) + L_f f_w, \\ \dot{z}(t) = A_z z(t) + B_z U(t) + L_z e(t), \\ \dot{f}_d(t) = L_e e(t), \\ \dot{e}(t) = L_f^{\mathrm{T}}[z(t) - \hat{z}(t)] + L_w e(t). \end{cases} \quad (23)$$

where

$$x(t)=\begin{bmatrix} x_{11} & x_{12} & x_{21} & x_{22} \end{bmatrix}^{\mathrm{T}}, z(t)=\begin{bmatrix} z_{11} & z_{12} & z_{21} & z_{22} \end{bmatrix}^{\mathrm{T}}, f_d=\begin{bmatrix} f_1 & f_2 \end{bmatrix}^{\mathrm{T}}, f_d=\begin{bmatrix} z_{13} & z_{23} \end{bmatrix}^{\mathrm{T}}, e=\begin{bmatrix} e_1 & e_2 \end{bmatrix}^{\mathrm{T}},$$

$$A_z=\begin{bmatrix} 0&1&0&0\\0&0&0&0\\0&0&0&1\\0&0&0&0 \end{bmatrix}, B_z=\begin{bmatrix} 0\\b_1\\0\\b_2 \end{bmatrix}, L_f=\begin{bmatrix} 0&0\\1&0\\0&0\\0&1 \end{bmatrix}, L_z=\begin{bmatrix} 0&0\\-\beta_{12}&0\\0&0\\0&-\beta_{22} \end{bmatrix}, L_e=\begin{bmatrix} -\beta_{13}&0\\0&-\beta_{23} \end{bmatrix}, L_w=\begin{bmatrix} -\beta_{11}&0\\0&-\beta_{22} \end{bmatrix}.$$

Define the augmented state vector $\zeta(t)=[x^{\mathrm{T}}(t), z^{\mathrm{T}}(t), f_d^{\mathrm{T}}(t), e^{\mathrm{T}}(t)]^{\mathrm{T}}$ and substitute the controller into the plant. The closed-loop dynamics can be written as

$$\dot{\zeta}(t) = \mathbb{A}\zeta(t) + \mathbb{B}_w f_w(\zeta(t)) + \mathbb{B}\nabla L(\zeta(t)). \quad (24)$$

with

$$\mathbb{A}=\begin{bmatrix} A_z & B_z K & L_f & 0 \\ 0 & A_z + B_z K & 0 & L_z \\ 0 & 0 & 0 & L_e \\ L_f^{\mathrm{T}} & -L_f^{\mathrm{T}} & 0 & 0 \end{bmatrix},$$

$$\mathbb{B}_w = [L_f \quad 0 \quad 0 \quad 0]^{\mathrm{T}},$$

$$\mathbb{B} = [-B_z K_3 \quad -B_z K_3 \quad 0 \quad 0]^{\mathrm{T}}.$$

We now establish stability of the closed-loop system.

**Theorem 3.2**

Suppose the nonlinear external disturbance $f_w$ and the gradient $\nabla L(\xi(t))$ of the objective function are continuously differentiable and bounded, and satisfy the sector bounds

$$f_w^{\mathrm{T}} f_w \le \zeta^{\mathrm{T}} M \xi$$
$$\nabla L^{\mathrm{T}} \nabla L \le \zeta^{\mathrm{T}} N \xi$$

where *M* and *N* are positive-definite diagonal matrices characterizing the nonlinearities. If there exists a symmetric positive-definite matrix *P* such that:

$$\Gamma = \begin{bmatrix} \vartheta & 2P \\ * & -I \end{bmatrix} < 0$$
$$\vartheta = P\mathbb{A} + \mathbb{A}^{\mathrm{T}} P + M + \mathbb{B}^{\mathrm{T}} PP\mathbb{B} IN + I$$

then the closed-loop system is stable.

**Proof:**

Consider the Lyapunov candidate:

$$V(\zeta(t)) = \zeta^{\mathrm{T}}(t) P \zeta(t)$$

Differentiating the above Lyapunov function yields:

$$\dot{V}(\zeta(t)) = \zeta^{\mathrm{T}}(t)(P\mathbb{A} + \mathbb{A}^{\mathrm{T}} P)\zeta(t) + 2\zeta^{\mathrm{T}}(t) P\mathbb{B}_w f_w(\zeta(t)) + 2\zeta^{\mathrm{T}}(t) P \nabla L(\zeta(t))$$
$$< \zeta^{\mathrm{T}}(t)(P\mathbb{A} + \mathbb{A}^{\mathrm{T}} P)\zeta(t) + 2\zeta^{\mathrm{T}}(t) P f_w(\zeta(t)) + 2\zeta^{\mathrm{T}}(t) P\mathbb{B} \nabla L(\zeta(t)) - f_w^{\mathrm{T}} f_w + \zeta^{\mathrm{T}}(t) M \zeta(t)$$

By Young's inequality we have:

$$2\zeta^{\mathrm{T}}(t) P\mathbb{B} \nabla L(\zeta(t)) \le \nabla L^{\mathrm{T}}(\zeta(t)) \mathbb{B}^{\mathrm{T}} PP\mathbb{B} \nabla L(\zeta(t)) + \zeta^{\mathrm{T}}(t)\zeta(t) \le \zeta^{\mathrm{T}}(t) \mathbb{B}^{\mathrm{T}} PP\mathbb{B} IN \zeta(t) + \zeta^{\mathrm{T}}(t)\zeta(t)$$

Therefore,

$$\dot{V}(\zeta(t)) < \varepsilon^{\mathrm{T}}(t) \Gamma \varepsilon(t)$$
$$\varepsilon(t) = [\zeta(t), f_w]$$

Since $\Gamma < 0$, it follows that $\dot{V}(\zeta(t)) < 0$. Scaling the above Young inequality gives:

$$\varepsilon^{\mathrm{T}}(t)(-\Gamma)\varepsilon(t) \ge \|\varepsilon(t)\|_2^2 \lambda_{\min}(-\Gamma) > \|\zeta(t)\|_2^2 \lambda_{\min}(-\Gamma).$$

Hence,

$$\dot{V}(\zeta(t)) < \varepsilon^{\mathrm{T}}(t)\Gamma\varepsilon(t) < \|\zeta(t)\|_2^2 \lambda_{\max}(\Gamma) < -\frac{\lambda_{\max}(-\Gamma)}{\lambda_{\max}(P)} V(\zeta(t)) < 0.$$

This indicates the closed-loop stability.

**Remark:**

The primary disturbance arises from engine exhaust flow, whose maximum rate is limited by the pipeline and cannot change abruptly. Thus, the total nonlinear disturbance is bounded and differentiable. Valve opening speed is mechanically constrained, so manifold pressure varies continuously. Therefore, the gradient of the augmented objective is also continuous. Since the exponential penalty function is differentiable and its maximum rate of change is bounded by pressure and pressure-ratio dynamics under valve limits, all assumptions of the theorem are satisfied.

# 5 Numerical Simulation and Validation

## 5.1 Simulation Verification Platform Setup

Based on the developed model and controller, a hardware-in-the-loop (HIL) simulation platform is built using a real-time simulator, a programmable logic controller (PLC), and MATLAB/Simulink (see Fig. 6). The platform consists of three parts: the simulator, the PLC, and the host PC. The host PC provides a human–machine interface for entering control targets and system parameters. The simulator runs the complete intake system model, including the control-valve model, the chamber temperature–pressure differential model, and the boundary-condition model. The PLC implements the control algorithms, including the external-penalty-function optimization and the ADRC algorithm.

In the simulation, Gaussian noise was added to the two chamber output signals to emulate sensor disturbances in a real test environment: the temperature signal received noise in the range (–0.1, 0.1), and the pressure signal noise in the range (–2, 2). At the same time, the nonlinear function shown in Fig. 4 was implemented as a 0.15 s delay element.

These simulations verify the intake-pressure control system's tracking performance under the proposed controllers and its disturbance-rejection capability during abrupt flow changes.

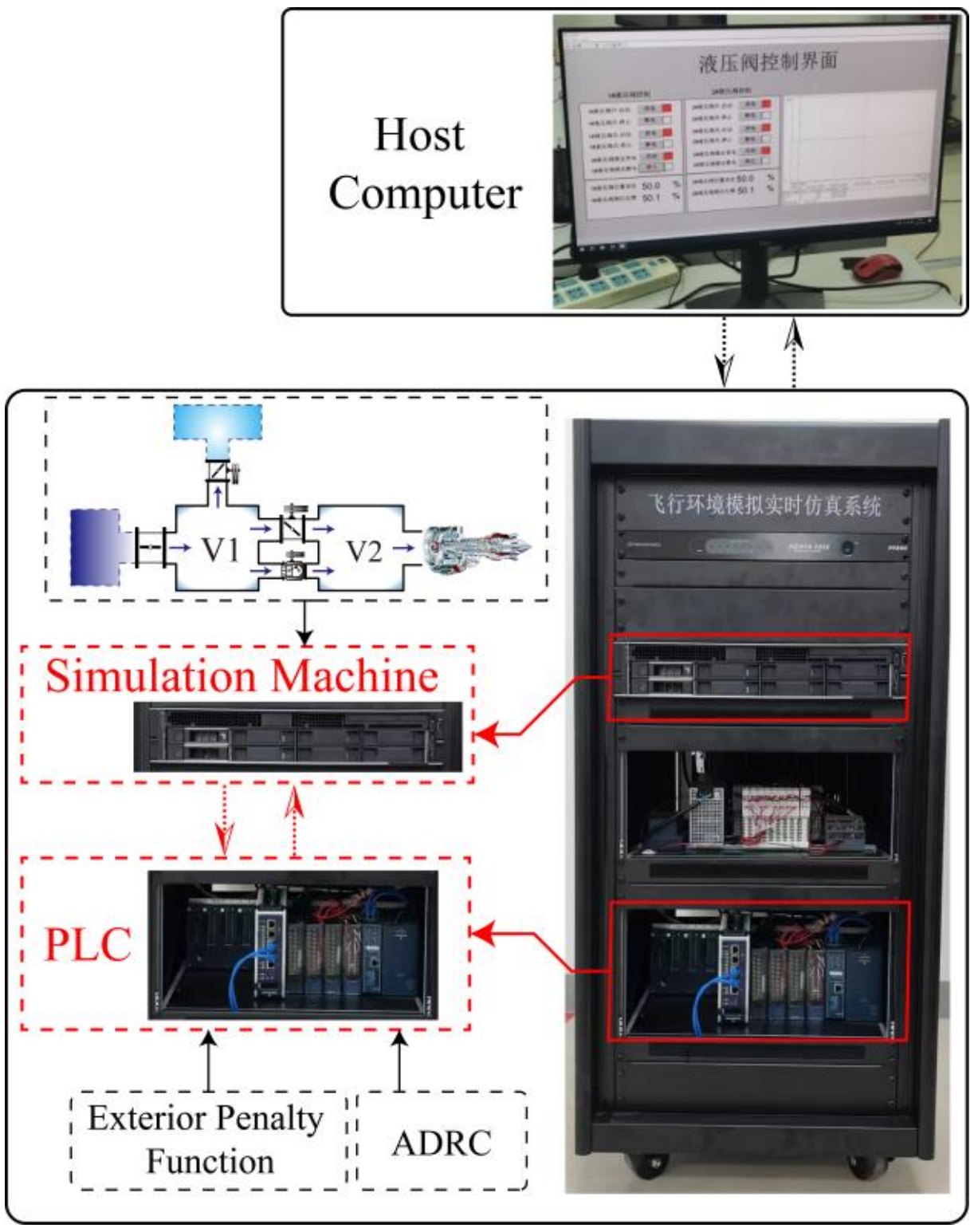

Fig. 6 Schematic diagram of the HIL simulation validation platform.

## 5.2 Simulation Testing and Validation

To evaluate the performance of the proposed coordinated external-penalty ADRC control method, a joint simulation of the intake system with two chambers and multiple valves was carried out. The inlet pressure and temperature of supply chamber V1 were set to 150 kPa and 253 K, respectively; the outlet pressure and temperature at V1's exhaust control valve were set to 101.325 kPa and 288 K. The equivalent volumes of chambers V1 and V2 were specified as 300 $m^3$ and 800 $m^3$, respectively. The time constants of valves Valve1, Valve2, and Valve_air were set to 2.5 s, 1.5 s, and 2.5 s, with diameters of 2.6 m, 1.2 m, and 2.6 m, respectively.

Based on experimental requirements, desired pressure profiles and engine extraction flow rates were defined (see Fig. 7) to assess control performance under various operating conditions. The control system composed of three independent PID loops served as the baseline. Comparative test results for the two control frameworks are shown in Fig. 8. The control system parameters are listed in Table 1.

Table 1 Main parameters of the controller

| Parameter | Value | Parameter | Value |
|---|---|---|---|
| $\gamma$ | 0.5 | $\mu$ | 0.001 |
| $\sigma$ | 0.01 | $l$ | 0.2 |
| $\omega_1$ | 2 | $\omega_2$ | 5 |
| $k_{11}$ | 0.001 | $k_{12}$ | 0.0001 |
| $k_{13}$ | 0.1 | $k_{21}$ | 0.04 |
| $k_{22}$ | 0.04 | $k_{23}$ | 1 |

The penalty factor in the table is denoted by $\gamma$. Its value was chosen as 0.5 to reflect the relevant orders of magnitude. The parameters $\mu$ and $\sigma$ tune the penalty strength; since chamber V2's pressure control is given higher priority, $\sigma$ carries the larger weight. The learning rate $l$ was set to 0.2. The ESO bandwidths were chosen as 2 Hz and 5 Hz, respectively, to avoid excessive tuning difficulties in practice.

The control gains for the two chambers are $k_{11}, k_{12}, k_{13}$ for V1, and $k_{21}, k_{22}, k_{23}$ for V2, where each triplet corresponds to Kp, Kd and the coordination gain. The coordination gain was set proportionally to the disturbance level each chamber experiences.

Three test phases were simulated:

1. **0–100 s (Uniform Flow Variation):** Engine flow decreases smoothly from 780 kg/s to 370 kg/s over the first 60 s, then remains constant until 100 s. During this phase, V1's setpoint is maintained at 65 kPa.
2. **100–250 s (Pressure Tracking):** V1's setpoint ramps from 65 kPa to 70 kPa at 125–150 s, holds until 155 s; then ramps to 75 kPa at 155–165 s, holds until 220 s; and finally returns to 65 kPa at 220–250 s. Engine flow varies according to the engine model[22].
3. **250–300 s (Disturbance Rejection):** Engine flow jumps from 280 kg/s to 550 kg/s over 265–270 s, then back to 280 kg/s over 280–285 s, with a peak rate of change of 180 kg/s$^2$. V1's setpoint remains at 65 kPa.

Throughout all tests, V2's pressure setpoint was fixed at 130 kPa. As shown in Fig. 8, the coordinated external-penalty ADRC delivers markedly better control quality than three independent PID loops, especially in the 250–300 s disturbance phase. The pressure-error curves under both control schemes appear in Fig. 9. Quantitative evaluation by root-mean-square error (RMSE) and maximum absolute error is summarized in Table 2.

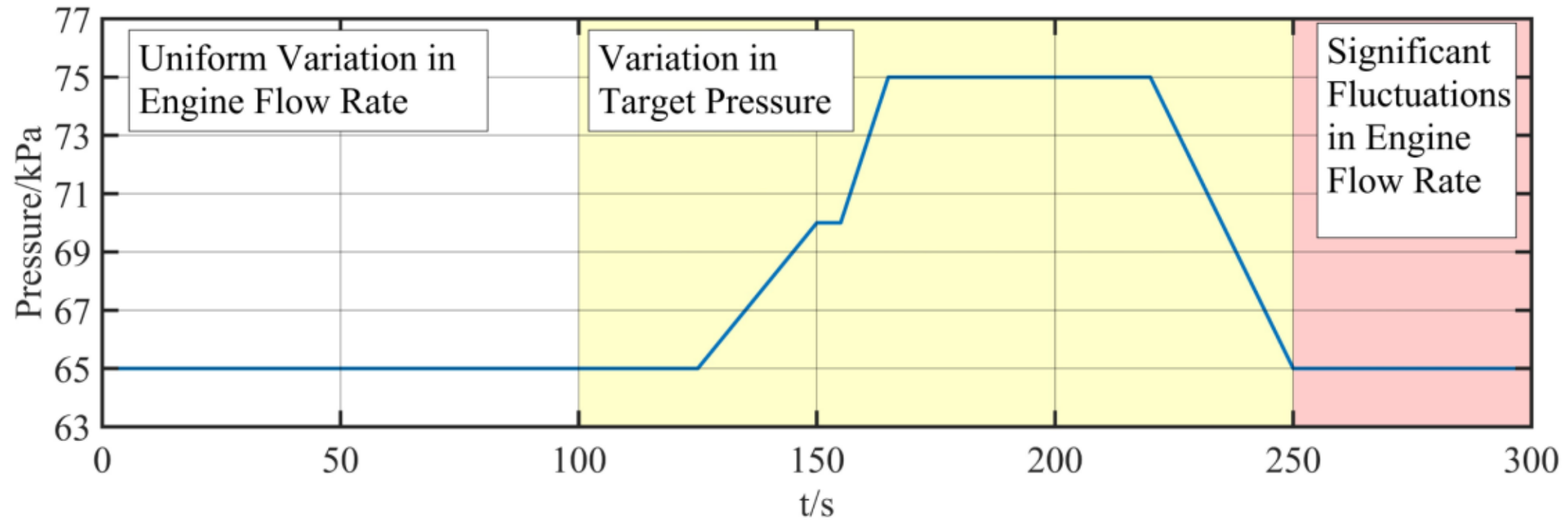

(a) Pressure control target

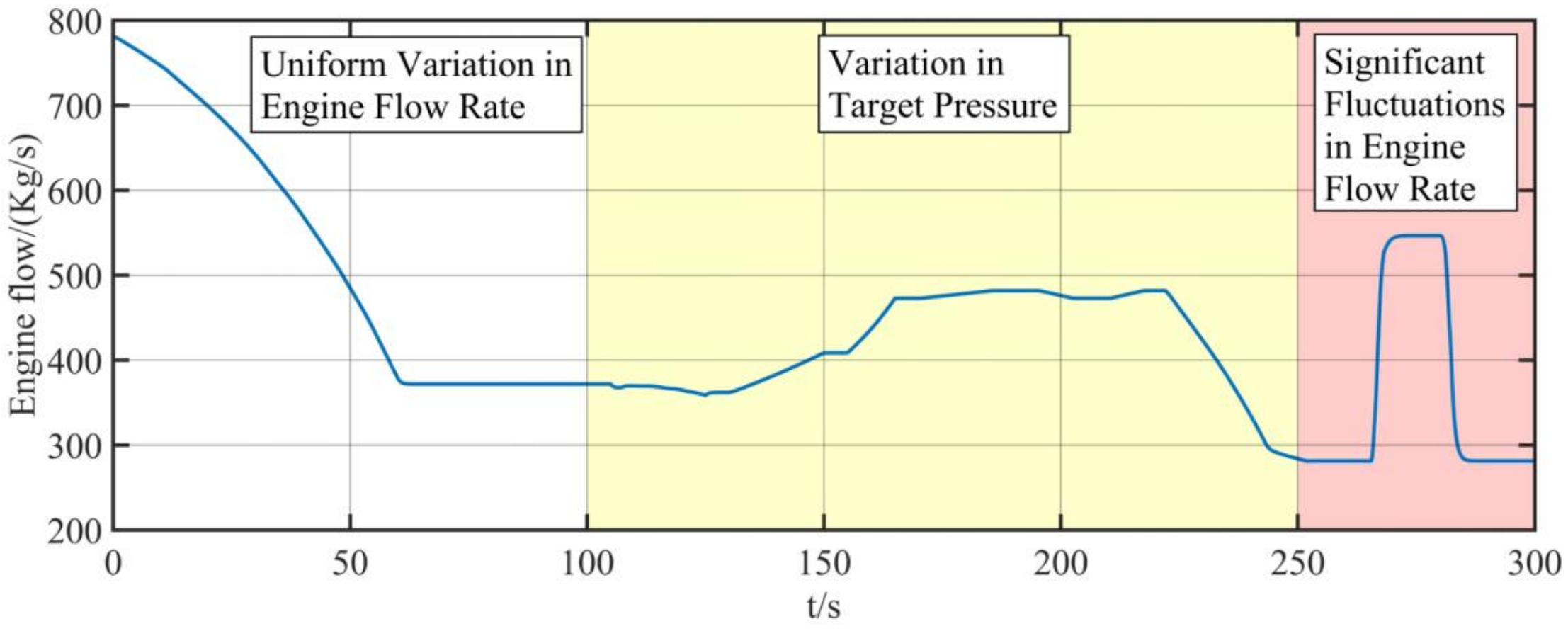

(b) Engine flow rate

Fig. 7 Test Condition Configuration

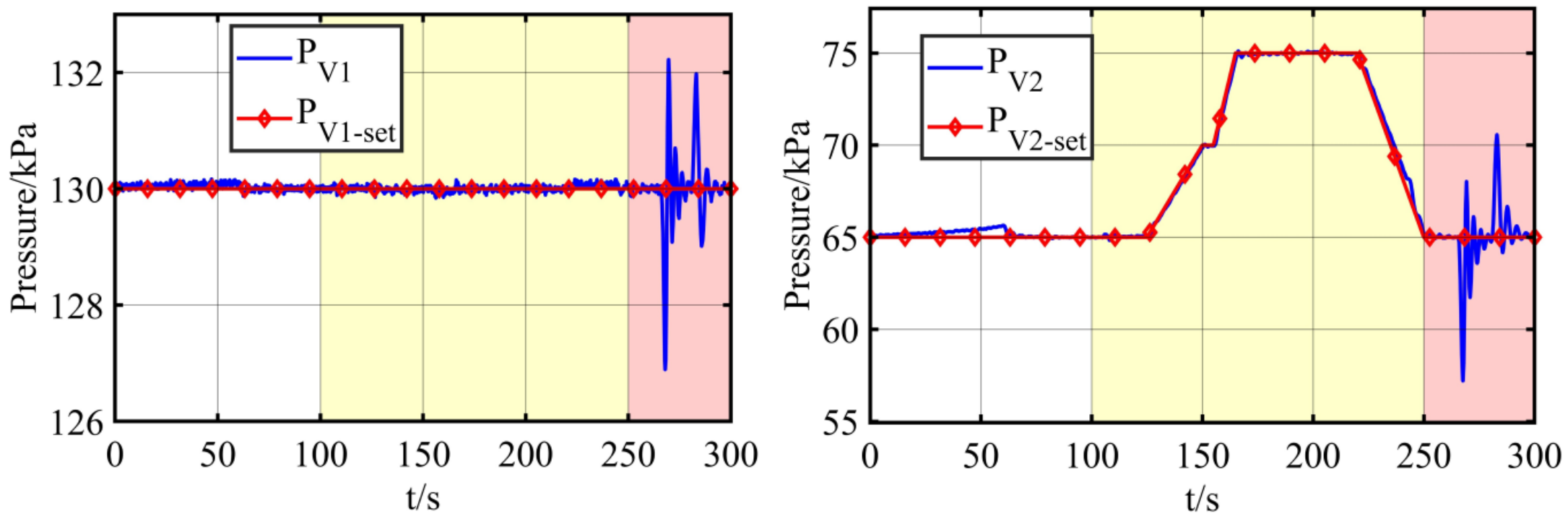

(a) Pressure Control Results for V1 and V2 under the PID Control Algorithm

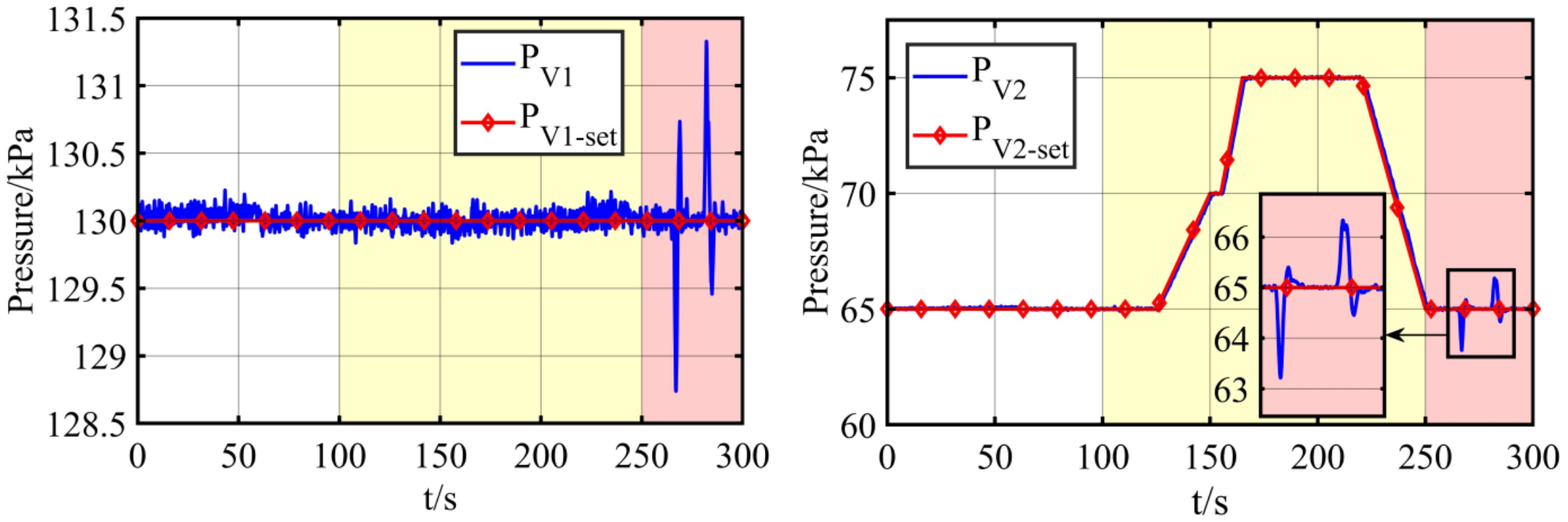

(b) Pressure Control Results for V1 and V2 under the Coordinated Control Algorithm

Fig. 8 Verification results

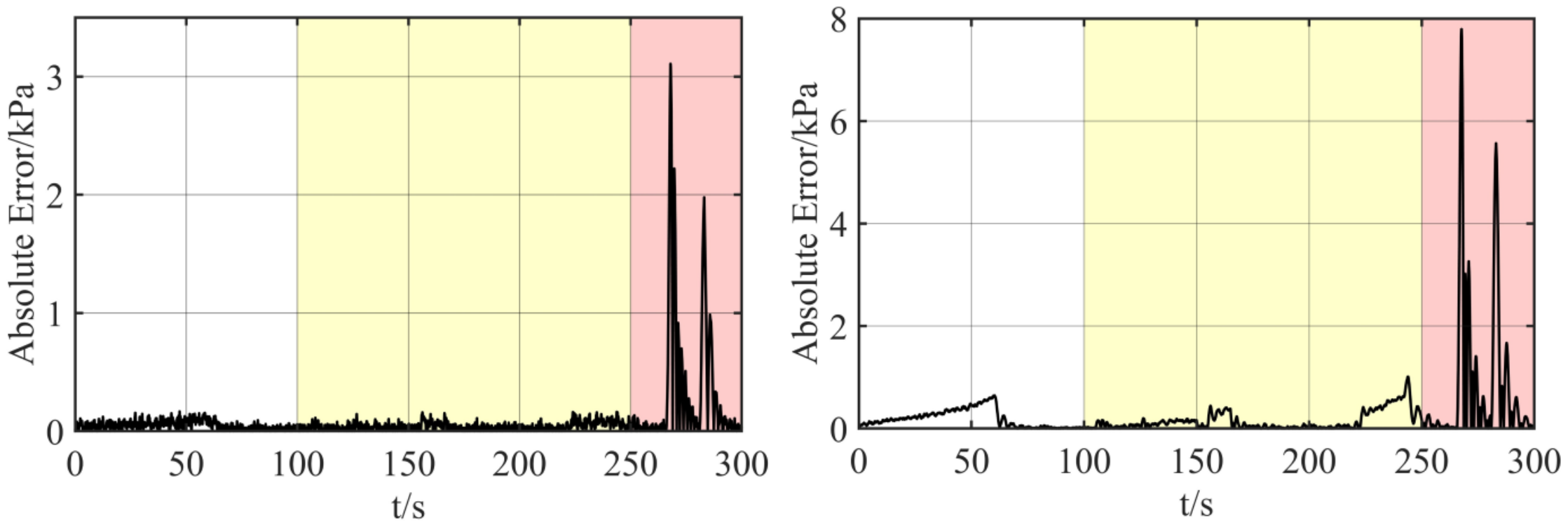

(a)Pressure Control Error of V1 using the PID Control Algorithm

(b)Pressure Control Error of V2 using the PID Control Algorithm

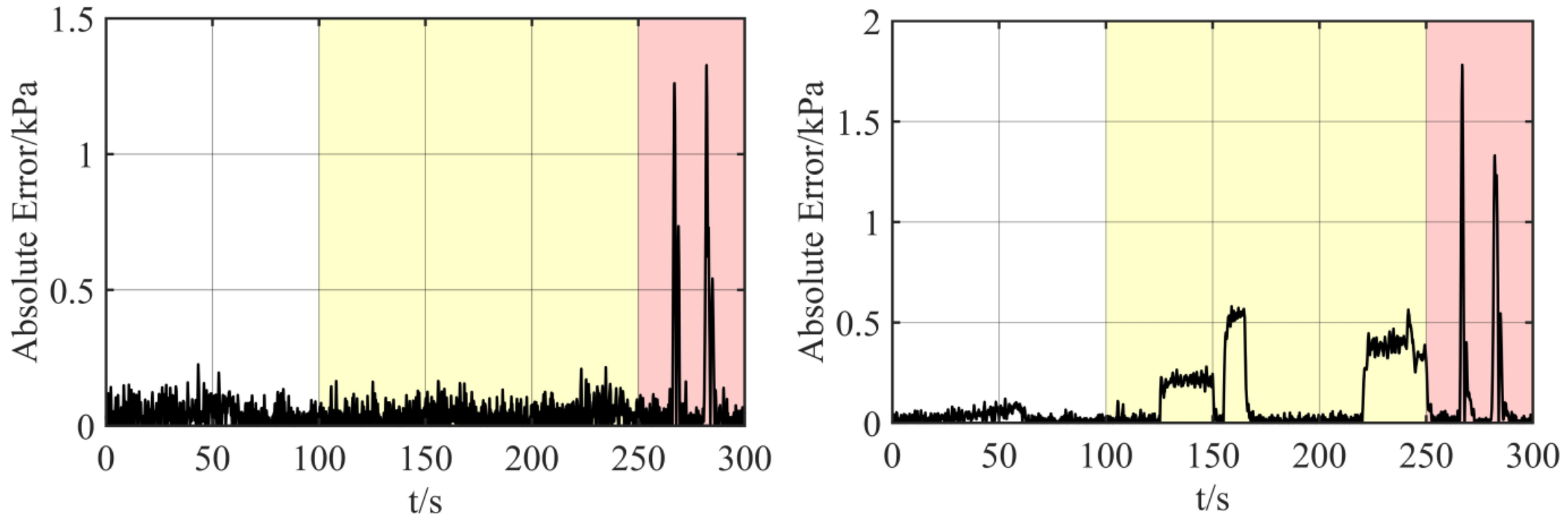

(c)Pressure Control Error of V1 using the Coordinated Control Algorithm

(d)Pressure Control Error of V2 using the Coordinated Control Algorithm

Fig. 9 Pressure control errors

Table 2 Control performance index

| | RSME_V1 | RSME_V2 | abs(err)max_V1 | abs(err)max_V2 |
|---|---|---|---|---|
| XT | 0.123 | 0.218 | 1.327kpa | 1.782kPa |
| PID | 0.278 | 0.736 | 3.107kPa | 7.793kPa |

Table 2 shows that the independent PID controllers fail to keep both chamber pressures within ±3 kPa during the tests. Chamber V2's peak error reaches 7.793 kPa. In contrast, the external-penalty coordinated ADRC algorithm consistently enforces the ±3 kPa constraint.

Valve-angle trajectories under both schemes are plotted in Fig. 10(a) for PID and Fig. 10(b) for ADRC. During the 250–300 s disturbance, the PID-based system exhibits large oscillations in all three valves—peak-to-peak fluctuations up to 27%—and strong cross-coupling, which induce coupled pressure oscillations in V1 and V2 (see Fig. 8(a)). In contrast, the coordinated ADRC produces synchronized valve movements with minimal oscillation (peak-to-peak < 5%), stable responses, and approximately 50% faster settling. This yields rapid convergence and effectively decouples the two chamber pressures.

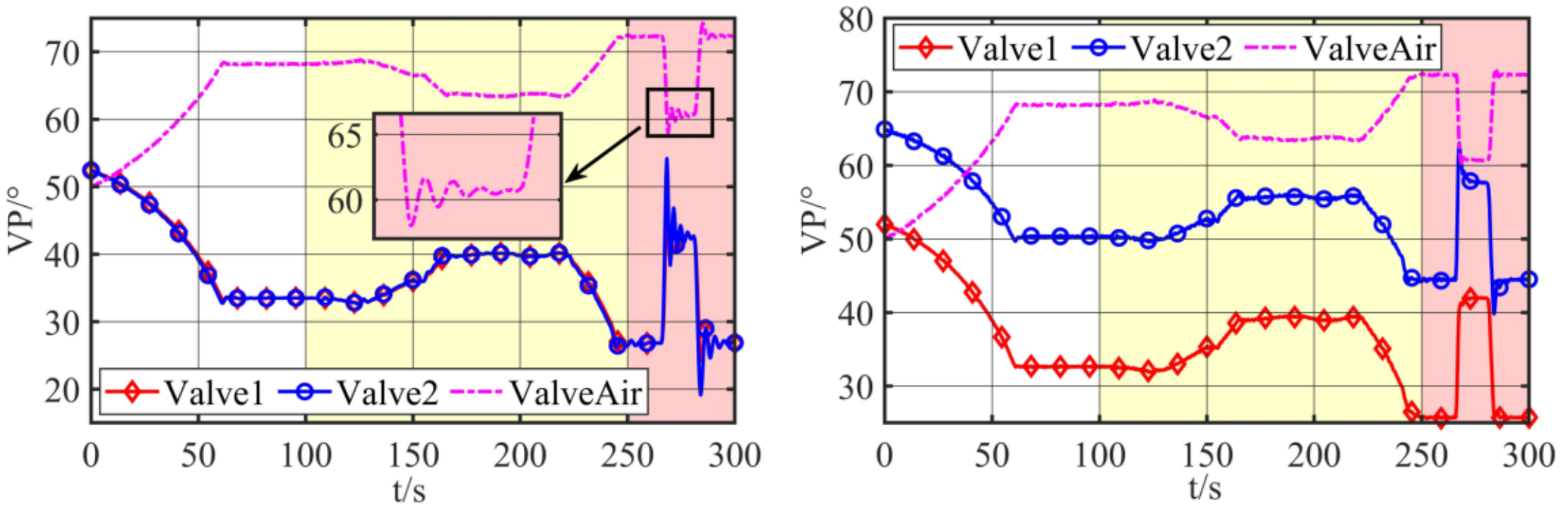

(a) Valve Opening Variation under PID-Based Control

(b) Valve Opening Variation under Cooperative Control

Fig. 10 Valve Position Variations

## 6 Conclusions

This paper addressed the challenges of multi-valve decoupling and strong-disturbance rejection in a high-altitude, multi-chamber intake system during engine transient tests. We proposed a coordinated ADRC method based on an external penalty function and validated it through theoretical analysis and hardware-in-the-loop

(HIL) simulation. The main conclusions are:

**1) Pressure-Error Constraint via External Penalty.**

The chamber-pressure safety limit was reformulated as an inequality-constrained optimization problem. An exponential penalty function, together with a gradient-based optimization algorithm, enables dynamic relaxation of constraints and provides a feasible domain for multi-objective coordination. The algorithm's global convergence was rigorously proven.

**2) Multi-Valve Coordinated ADRC.**

We enhanced the existing distributed ADRC framework by integrating a coordination-optimization term with the disturbance-compensation mechanism. This yielded a multi-valve coordinated ADRC controller that overcomes the valve-saturation and pressure-oscillation issues of traditional methods. Lyapunov stability theory guarantees the asymptotic stability of the closed-loop system.

**3) HIL Simulation Results.**

Tests on a MATLAB/Simulink + PLC HIL platform show that the proposed method confines the V2 chamber's pressure error within 3 kPa (maximum 1.782 kPa), a 77.1 % reduction compared to PID control. Under a flow disturbance rate of 180 kg/s$^2$, valve-opening oscillations shrink by 81.5 % (±5 % vs. ±27 %), confirming the method's disturbance-rejection and decoupling capabilities.

This study has focused on pressure coordination in a dual-chamber intake system. Future work will introduce temperature parameters to develop a coupled pressure–temperature control model and investigate coordinated temperature–pressure control under high and low temperature transient conditions. This will enhance environmental simulation accuracy and overall stability in complex thermo-coupled systems.

2007;22(12):2134–2138 [Chinese].